\documentstyle [12pt] {article}
\begin{document}
\begin{flushright}
PRL-TH-94/08\\
March 1994
\end{flushright}
\vskip .5in
\begin{center}
{\Large \bf Degenerate Neutrinos in Left Right}\\
 {\Large \bf Symmetric Theory}\\[50pt]
{\bf Anjan S. Joshipura}\\ \vskip .2in
 Theory Group, Physical Research Laboratory\\Navrangpura,
Ahmedabad 380009, India\\[40pt]
\end{center}
\begin{abstract}
Various hints on the neutrino masses namely, ({\em i})
the solar neutrino deficit ({\em ii}) the atmospheric neutrino
deficit ({\em iii}) the need for the dark matter
 and/or ({\em iv}) the
non-zero neutrinoless double beta decay collectively imply that
all the three neutrinos must be nearlty degenerate. This feature can be
understood in the left right symmetric theory. We
 present a model based on the group $SU(2)_{L}\times
SU(2)_R\times U(1)_{B-L}\times SU(2)_H$ which can explain the
required departures from degeneracy in neutrino masses
and large
mixing among them  without assuming any of the mixing in the
quark or charged lepton sector to be large as would be expected
in a typical $SO(10)$ model.
\end{abstract}
\newpage
No laboratory experiment has unambiguously detected the mass for
the neutrino so far. But there exists variety of hints
 \cite{smir} which
when taken in totality \cite{moh,asj}  are strong enough to suggest a definite
pattern for the masses of the known neutrinos. These hints come
from ({\em i}) deficit in the solar neutrinos \cite{four}
({\em ii}) deficit in the
low energy atmospheric neutrinos \cite{kam2} ({\em iii}) need for about
 30\% hot
dark matter\cite{dm} and ({\em iv}) indications that neutrinoless double beta
decay may actually be taking place \cite{moe}. These observations when
attributed to neutrino masses put strong restrictions on the
neutrino (mass)$^2$ differences $\Delta$ and mixing $\sin^2 2
\theta$ as well as the absolute values of their masses
\cite{smir}.
It was shown in
ref. \cite{moh,asj} that any of the observation ({\em iii}) or
({\em iv}) when combined with
({\em i}) and ({\em ii}) imply that all the neutrinos must be nearly
degenerate in mass if there are only three light neutrinos.

The near degeneracy of the neutrino masses is very
different from the hierarchy observed in the masses of other
fermions. But this seemingly different pattern can be naturally
incorporated  \cite{moh,asj} into the seesaw mechanism for
 the neutrino mass generation.
This framework (when suitably augmented by a horizontal symmetry
) is capable of explaining not only the near degeneracy in the
neutrino masses but it can also lead  \cite{asj} to the observed departures
from the degeneracy. Specifically, one expects \cite{asj}
\begin{eqnarray}
m_{\nu_1}=m_0-m_u^2/M,\nonumber\\
m_{\nu_2}=m_0-m_c^2/M,\nonumber\\
m_{\nu_3}=m_0-m_t^2/M
\end{eqnarray}
where $m_0$ is the universal mass for the light neutrino while $M$
represents the large majorana mass for the right handed neutrinos.
The above equation leads to
\begin{equation}
\frac{|\Delta_{21}|}{|\Delta_{32}|}\approx
\left(\frac{m_c}{m_t}\right)^2\approx (1-3)\times  10^{-4}
\end{equation}
This nicely reproduces the hierarchy required to simultaneously
solve the solar and atmospheric neutrino problem. Moreover, if
$m_0$ is $\sim$2 eV as required for the hot dark matter \cite{dm} or for
obtaining the neutrinoless double beta decay  \cite{smir} at the present
experimental level, and if $M$ is identified with  the grand unification
scale ($\sim 10^{16}$ GeV), one obtains  \cite{asj}  $\Delta_{12}$ in the
range required for the solution of the solar neutrino problem
through the MSW \cite{msw} mechanism.

The neutrino sector seems
to be distinguished from other fermions also in respect of mixing
among them. For example,
the atmospheric
neutrino problem \cite{smir}  can be solved only if $\nu_{\mu}$
is strongly mixed with the other neutrino say $\nu_{\tau}$. Typically,
 $\sin^2\;2 \theta_{\mu\tau}\sim .5$. The solution of the solar neutrino
problem through the vacuum oscillations  also needs large mixing
between $\nu_e$ and say $\nu_{\mu}$. This is quite different
from the quark sector where all the mixing angles are known to
be small. One would like to understand this feature of the
neutrino sector along with their almost degeneracy.
We present a possible way to understand this in the context of
the left right symmetric theory and discuss it through  an explicit model.
\footnote{Various models have been recently proposed
 \cite{cliff,valle,moh1,moh2}
while the present work was in the progress. We shall make a
comparison of these models with the present one at the end.}

The structure of the neutrino masses given in eq.(1) can arise
form the following seesaw mass matrix:
\begin{equation}
M_{\nu}=\left( \begin{array}{cc}
          m_0\;I&M_{\nu D}\\
          M_{\nu D}^T&M\;I \end{array} \right)
\end{equation}
The above structure arises naturally \cite{sanj} in a left right symmetric
theory with an extra discrete parity $D$ which connects the left
and the right-handed sectors.
 Because of this symmetry, the breakdown of the $SU(2)_R$ at a
high scale naturally induces a small  vacuum expectation value
(vev) for the left
handed triplet Higgs field and thus leads to a non-vanishing
contribution to the majorana masses of the left
handed neutrinos.  These dominate \cite{sanj} over the
conventional seesaw contribution for natural values of parameters.
Hence, if some horizontal symmetry makes these masses identical,
the physical neutrino masses would be almost degenerate. The
conventional seesaw contribution then would lead to departures
form this degeneracy with the structure similar to the one
displayed in eq.(1).

While the basic scenario outlined above
 follows \cite{moh,asj} from  simple
considerations, the details require the presence of a
complicated underlying structure specifically in the Higgs
sector. A candidate model was first proposed in ref. \cite{asj}
 which
used the horizontal $SU(2)$ symmetry. The present one as well as
some of the recent models \cite{valle,moh1} are also based on
this symmetry. Our basic aim here is to understand both the
departures of  neutrino masses from degeneracy as well
difference in mixing pattern between neutrinos and quarks in a
qualitatively different manner compared to ones presented
 in \cite{asj,valle,moh1}.

In the context of a typical seesaw model, one expects
 \cite{lang} relations
not only between the masses of neutrinos and other fermions but
also between their mixings. As a result, the large mixing is not
natural in this case.
There one expects  $M_{\nu D}$ in eq.(3) to be similar to a typical
 up quark
mass matrix and $M_{down}$ to $M_{leptons}$. The  mixing in the
neutrino sector is then related to the quark sector and hence would
be expected to be small. This can be avoided if the majorana
mass matrix for the right handed neutrinos has some texture
 \cite{smir1}.
This possibility does not exist in the present case because of
the assumed left right symmetry. This would automatically make
the right handed majorana mass matrix  proportional to the left
handed majorana mass matrix. The latter is required to be
proportional to identity (see eq.(3)) if one wants to have
degenerate masses for the neutrinos. We propose a way out
which naturally leads to differences in the mixing pattern
between quarks and neutrinos without giving up the basic left
right symmetry.

We retain the underlying left right symmetric framework which
naturally explains the degeneracy and work for simplicity with
the gauge group
 $G_{LR}\equiv SU(2)_L\times SU(2)_R\times U(1)_{B-L}$.
An $SU(2)_H$ symmetry is  introduced to obtain the degenerate
neutrinos and a $G_{LR}\times SU(2)_H$ singlet fermion $N$ is
introduced to obtain different mixing pattern among neutrinos
compared to other fermions. The $SU(2)_H$ group can either be
a softly broken global symmetry or could be gauged. Fermionic
generations are taken to transform as triplets under $SU(2)_H$
and the Higgs fields
transform as follows under  $SU(2)_L\times SU(2)_R\times SU(2)_H$:
\begin{eqnarray}
\Phi_{ab}\sim (2,2,5)\;\;\;\;\;\;\Phi\sim (2,2,1) \nonumber\\
\Delta_L\sim (3,1,1)\;\;\;\;\;\;\Delta_R \sim (1,3,1) \nonumber\\
\phi_L\sim (2,1,3)\;\;\;\;\;\;\phi_R \sim (1,2,3).
\end{eqnarray}
The Yukawa couplings of the quarks ($Q_{L,R}$) and leptons ($l_{L,R}$)
are given by
\begin{eqnarray}
{\cal L}_{Y}&=&h_q\;\overline{Q}_{aL}\Phi
Q_{aR}+h_q'\;\overline{Q}_{aL}\hat{\Phi} Q_{aR} +
\gamma_q\;\overline{Q}_{aL}\Phi_{ab}Q_{bR}
+\gamma_q'\;\overline{Q}_{aL}\hat{\Phi}_{ab}Q_{bR} \nonumber\\
& &+h_l\;\overline{l}_{aL}\Phi
l_{aR}+h_l'\;\overline{l}_{aL}\hat{\Phi} l_{aR} +
\gamma_l\;\overline{l}_{aL}\Phi_{ab}l_{bR}
+\gamma_l'\;\overline{l}_{aL}\hat{\Phi}_{ab}l_{bR} \nonumber \\
& &+f \left[ l_{aL}^T C{\bf
\epsilon\tau.\Delta_L}l_{aL}+L\leftrightarrow R\right]
+h_N\left[\overline{l}_{aL}\phi_{aL}+L \leftrightarrow R \right]
\;N+\frac{1}{2}M\;N^T\;CN
\end{eqnarray}

Where $a,b=1,2,3$ are the generation indices;
$\hat{\Phi}\equiv \tau_2\Phi^*\tau_2$.
Both the $\Phi$ and $\Phi_{ab}$ contain two neutral fields which
could acquire vacuum expectation value (vev). We shall denote by
$\kappa,\kappa'(\kappa_{ab},\kappa_{ab}')$ the vev of the
neutral components contained in $\Phi (\Phi_{ab})$. The mass
matrices for the charged fermions ($f=U,D,E$) are then given by
\begin{eqnarray}
(M_U)_{ab}&=&(h_q\kappa+h_q'\;\kappa')\delta_{ab}
          +\gamma_q\;\kappa_{ab}+\gamma_q'\;\kappa'_{ab} \nonumber\\
(M_D)_{ab}&=&(h_q\kappa'+h_q'\;\kappa)\delta_{ab}
          +\gamma_q\;\kappa'_{ab}+\gamma_q'\;\kappa_{ab} \nonumber\\
(M_E)_{ab}&=&(h_l\kappa'+h_l'\;\kappa)\delta_{ab}
           +\gamma_l\;\kappa'_{ab}+\gamma_l'\;\kappa_{ab}
\end{eqnarray}
The structure of the neutrino masses is more complicated. Assume
that the $SU(2)_R$ symmetry is broken by the large vev of
$\Delta_R$ and $\phi_{R}$. The  vev of the
right handed triplet automatically induces the vev for the left
handed triplet \cite{sanj} if the potential is to respect
$G_{LR}\times D$. This follows from the following types of terms
in the Higgs potential
\begin{equation}
V_{\Delta}=\mu^2\;Tr(\Delta_L^\dagger \Delta_L
           +\Delta_R^\dagger \Delta_R)
           +\lambda\;Tr\left((\Delta_L^\dagger \Delta_L)^2+
(\Delta_R^\dagger \Delta_R)^2\right)
           +\delta \;
Tr\left(\Delta_L^\dagger\Phi\Delta_R\Phi^\dagger\right)+....
\end{equation}

Where we have retained only typical terms which lead to the
following relation at the minimum (more general analysis can be
found in \cite{sanj})
\begin{equation}
<\Delta_L><\Delta_R>\approx \gamma \kappa^2
\end{equation}
where, $\gamma$ is related to the parameters in $V_{\Delta}$ and
$\kappa$ is a typical $SU(2)_L$ breaking vev of the field $\Phi$.
Very similar hierarchy also exists among the vev of the fields
 $\phi_{L,R}$.
This would follow  form the terms in the Higgs potential of the
following type:
\begin{equation}
V_{\phi}=\mu'^2(\phi_L^\dagger \phi_L+\phi_R^\dagger \phi_R)
+\lambda'\left((\phi_L^\dagger \phi_L)^2
+(\phi_R^\dagger \phi_R)^2\right)
           +\delta' \left(\phi_L^\dagger\Phi\phi_R\right)+....
\end{equation}
This leads to the following relation
\begin{equation}
<\phi_L><\phi_R>\approx \gamma' \kappa \delta'
\end{equation}
Hence if the vev for $\phi_R$ is required to be very large
as we will do in the following then the induced vev for $\phi_L$
will automatically be suppressed. This suppresses the mixing of the
left handed neutrinos with the field $N$  allowing at the same
time a large mixing between the right handed neutrinos and $N$.
If we neglect the former mixing then the neutrino mass
 matrix is given
in the basis $(\nu_L^c,\nu_R,N)$ by
\begin{equation}
\cal{M_{\nu}}=\left( \begin{array}{cc}
                      m_0\;I&\hat {M}_{\nu D}\\
                      \hat {M}_{\nu D}^T&\hat {M}_R\\
                      \end{array}  \right)
\end{equation}
We have the following form for various matrices:
\begin{equation}
\hat{M}_{\nu D}\approx\left(  \begin{array}{cc}
                         M_{\nu D}&\begin{array}{c}
                              0\\
                              0\\
                              0\\  \end{array}\\
                      \end{array}
              \right) \end{equation}
\begin{equation}
\hat{M_R}\approx\left(  \begin{array}{cccc}
             M_0&0&0&M_1\\
             0&M_0&0&M_2\\
             0&0&M_0&M_3\\
             M_1&M_2&M_3&M\\  \end{array} \right) \end{equation}
$M_a\equiv h_N<\phi_{Ra}>$; $M_{\nu D}$ is a $3\times 3$ matrix
 following from eq.(5):
\begin{equation} (M_{\nu D})_{ab}=(h_l\kappa
+h_l'\kappa')\delta_{ab}+\gamma_l\kappa_{ab}+\gamma_l'\kappa_{ab}'
\end{equation}
It follows from eqs(5) and (8) that
\begin{equation}
M_0=f<\Delta_R>\;\;\;\;\;\;m_0=f^2\frac{\gamma \kappa^2}{M_0}
\end{equation}

The effective masses of the three light neutrinos are given by
the matrix:
\begin{equation}
m_{eff.}\approx m_0 I-M_{\nu D}M_R^{-1}M_{\nu D}^T
\end{equation}
with
\begin{equation}
M_R^{-1}\approx\frac{1}{D}\left(  \begin{array}{ccc}
             D_1&M_1\;M_2\;M_0&M_1\;M_3\;M_0\\
             M_1\;M_2\;M_0&D_2&M_2\;M_3\;M_0\\
            M_1\;M_3\;M_0&M_2\;M_3\;M_0&D_3\\
             \end{array} \right) \end{equation}
where,
\begin{eqnarray}
D_1=M_0(M_0M-M_2^2-M_3^2) \nonumber \\
D_2=M_0(M_0M-M_1^2-M_3^2) \nonumber \\
D_3=M_0(M_0M-M_2^2-M_1^2) \nonumber \\
D=M_0^2(M_0M-M_2^2-M_3^2-M_1^2)
\end{eqnarray}
It follows from eqs.(16,17) that the terms induced by the coupling
between $N$ and the right handed neutrinos allow for a general
mixing among neutrinos even in the extreme case of the diagonal
$M_{\nu D}$.
This therefore allows us to understand the observed difference
in the mixing among neutrinos compared to  other fermions.
While this possibility can be realized in general, in the
following we discuss a specific  case which has the
merit of being economical.
In this example, the mixing in
the quark sector is correlated to that in the  leptonic sector.
Normally, such a situation would arise in typical models based
 on $SO(10)$.
The additional $SU(2)_H$ symmetry turns out be restrictive
in our case and a similar situation can be realized
even with the gauge group $G_{LR}$.
This happens if ({\em i})
all the primed Yukawa couplings in eq.(5) are set to zero. This
would be true in the supersymmetric theory or if one imposes
some softly broken Peccei Quinn symmetry \cite{babu}
 and ({\em ii}) The vev $\kappa_{ab}$ and
$\kappa'_{ab}$ are real. The $\kappa_{ab}$  can then be chosen
 diagonal by
a proper $SU(2)_H$ rotation. In this case, eqs. (6,14) can be used to
show that

\noindent (a) The $M_{\nu D}$ and $M_U$ are diagonal.

\noindent (b) The following relations hold among various masses:
\begin{eqnarray}
\frac{m_b-m_d}{m_{\tau}-m_e}&=&\frac{m_s-m_d}{m_{\mu}-m_e} \\
\frac{m_t-m_u}{m_{c}-m_u}&=&\frac{m_3-m_1}{m_{2}-m_1}
\end{eqnarray}
$m_{1,2,3}$ are the eigenvalues of $M_{\nu D}$.  Eq.(19)
relates the leptons and the quark masses and follows here without
using any grand unification. This is  is seen to be reasonably well
 satisfied. Note however that both these
relations would be expected to receive corrections if the group
$G_{LR}\times SU(2)_H$ is broken at a high scale.

\noindent (c) The matrices $M_D$ and $M_E$ can be diagonalized
by the same matrix which would coincide with the Kobayashi
Maskawa matrix in this case. \footnote{Note that with the assumption
 ({\em i}) and ({\em ii}) above,
the Kobayashi Maskawa matrix becomes real even if $h_q$ and
$\gamma_q$ are complex. Thus one has to look for CP violation
elsewhere or has to relax these assumptions. We shall not
 discuss these issue here as our main motivation is
the neutrino sector. }
Hence, the mixing among the charged leptons cannot be large and
we shall neglect it completely in the following.
The large mixing among neutrinos could come about
because of the presence of the additional singlet which allows
the $M_R^{-1}$ to have a texture and a general form given by eq.(17).
 The details will depend upon
various parameters entering the matrix $m_{eff.}$.  We discuss
below a specific choices for the ranges of parameters needed to
obtain the realistic pattern. We will assume $M_1$ to be very
small and set it to zero.  The neutrino masses then are given by
\begin{equation}
m_{eff.}\approx m_0\; I -\frac{1}{D}\left(  \begin{array}{ccc}
             m_1^2D_1&0&0\\
             0&m_2^2 D_2&M_0M_2M_3m_2m_3\\
             0&M_0M_2M_3m_2m_3&m_3^2D_3\\
             \end{array} \right) \end{equation}
The masses $m_{1,2,3}$ are restricted by eq.(20). This allows for a
non-hierarchical values of $m_{1,2,3}$ but we assume more
natural possibility of the hierarchy $m_1\;<<\;m_{2,3}$. In
this case, eq.(20) implies
\begin{equation}
\frac{m_c}{m_t}\approx\frac{m_2}{m_3}.
\end{equation}
Realistic pattern for the mixing and masses now require the hierarchy
$M_0\sim M\;<M_2\;<M_3$.
In this limit, the
mixing between the second and the third neutrino and the masses of the
neutrinos are given by
\begin{eqnarray}
m_{\nu1}\approx m_0-\frac{m_1^2}{M_0} \nonumber \\
m_{\nu2}\approx m_0-\frac{m_2^2\;M\cos^2\theta_{23}}{M_2^2} \nonumber \\
m_{\nu3}\approx m_0-\frac{m_2^2}{M_0 \sin^2\theta_{23}}
\nonumber \\
\tan^2\theta_{23}\approx(\frac{m_c}{m_t})^2\;(\frac{M_3}{M_2})^2
\end{eqnarray}
It follows from there that
\begin{equation}
\frac{|\Delta_{32}|}{|\Delta_{21}|}\approx
\frac{M_2^2}{M M_0\tan^2\theta_{23}}\;\;\;\;\;
\Delta_{21}\approx -2 m_0m_2^2\frac{M\;\cos^2\theta_{23}}{M_2^2}
\end{equation}
Typically, for
\begin{equation}
\frac{M_2}{M_3}\sim \;\;\;\; \frac{(M\;M_0)^{1/2}}{M_2}\sim \frac{1}{30}
\end{equation}
one obtains
\begin{eqnarray}
M_0\approx  10^{13} \hbox {GeV} \nonumber \\
\sin^22\;\theta_{23}\approx 0.5 \nonumber \\
\frac{|\Delta_{21}|}{|\Delta_{32}|}\approx
(\frac{m_c}{m_t})^2\sim 10^{-4}
\end{eqnarray}
if $\Delta_{21}\sim 10^{-6}\hbox {eV}^2 $ and $m_2\sim m_c$.
These values are in the right range needed to solve both the
atmospheric neutrino and the solar neutrino problem. Moreover,
with the $M_0$ given in eq.(26), the common mass  $m_0$ is
 fixed to be in the eV
range (see eq. (15) ) as required for solving the dark
 matter problem if $f^2\gamma\sim 1$.
It is seen from eq.(25) that if $M\sim M_0\sim 10^{13}$ GeV then
$M_3$ is fixed to be around the grand unification scale. Thus,
one needs to associate, two physically distinct scales with the
vev of $\Delta_R$ and $\phi_R$. These different scales then
lead to the hierarchy in the values of $\Delta_{21}$ and $\Delta_{23}$.
The $\nu_e-\nu_{\mu}$ mixing would depend on the KM matrix and
on $M_1$ both of which we have ignored in the above for
simplicity.  It should be possible to generate the desired
mixing when these parameters are kept non-zero
in view of the general structure (eq. (17)) possessed
by $m_{eff.}$

If we set $M_{1,2,3}$ zero in eq.(17) then, the additional singlet
$N$ decouples. This corresponds to the usual situation. In this
case, one could understand the large mixing among neutrinos if
either $M_E$ is unrelated to $M_D$ and thus could admit a large
mixing or the Dirac masses for the neutrinos are unrelated to
the up quark masses. While such relations are typical of
$SO(10)$ and are interesting from the point of view of economy,
they are not automatic and can be avoided in the $SO(10)$ model
by invoking more Higgs fields. This has been used to generate
large mixing in ref. \cite{valle,moh1} where the presence of
 a doublet vev coming
from the 126 dimensional representation lead to inequality of
$M_D$ and $M_E$. In this case, one could have a large mixing
among charged leptons without conflicting with the small mixing
of  the KM matrix.  Note that in our case also the $M_D$ and
$M_E$ could be unrelated if the primed couplings in eq.(6) are retained.
 In
such a case, one could accommodate the large mixing between neutrinos
without invoking any singlet fermions. But the singlet fermion
makes it possible to have large mixing even without sacrificing
the relationship between $M_D$($M_U$) and $M_E$($M_{\nu D})$.

The present model is a logical combination of the models in
ref \cite{asj} and \cite{cliff}. Bamert and Burgess used the
singlet in order to generate departures from degenerate neutrino
 spectrum using the conventional seesaw mechanism. In the
present case, the left right symmetry naturally explains the dominance
of the degenerate mass term over the family dependent contributions.
The singlet is used here mainly to decouple the mixing in the
neutrino sector from the rest of fermions.

In summary, we have discussed a specific model to generate the
almost degenerate spectrum for the neutrino masses and mixing.
The model presented here provides a concrete example of the
proposals in ref. \cite{moh,asj}. The salient feature of the
model is the left right symmetry and a gauge singlet fermion
which together is shown to lead to the  pattern of neutrino
masses and mixing desired from the point of view of solving the
solar, atmospheric and the dark matter problem simultaneously.
While we have worked here with the left right symmetric gauge
group $G_{LR}$, the required scales call for embedding of this
group into an $SO(10)$ type of grand unified model with an
intermediate scale around 10$^{13}$ GeV.

\noindent {\bf Acknowledgments}

I would like to thank participants of the Workshop on High
Energy Physics Phenomenology held at  Madras,
India- particularly- K.S. Babu, C. Burges
S. Rindani and J. Valle for numerous discussions related to
the present work and to J. Valle and C. Burges for informing me
about their work.


\newpage
\begin{thebibliography}{99}
\bibitem{smir} Recent review is contained in A. Yu. Smirnov,
ICTP preprint IC/93/388.
\bibitem{moh} D. Caldwell and R.N. Mohapatra, Phys. Rev. {\bf D
48} (1993) 3259.
\bibitem{asj} A. S. Joshipura, Physical Research Lab. Report,
 PRL-TH/93/20 (1993)
\bibitem{four} K. Lande {\it et al} in  Proc. XXVth Int. Conf. on
High Energy Physics, Singapore, ed. K.K. Phua and Y. Yamaguchi,
World Scientific (Singapore 1991);
K.S. Hirata {\it et al} Phys. Rev. Lett. {\bf 66}  (1991) 9;
P. Anselman {\it et al} Phys. Lett. {\bf B285}
(1992) 376;
A. I. Abazov {\it et al} Phys. Rev. Lett. {\bf
67}  (1991) 3332; V. Gavrin, talk at the Int. Conf. on High
Energy Physics, Dallas, 1992.
\bibitem{kam2} K. S. Hirata {\it et al} Phys. Lett. {\bf B 280}
(1992) 146;
 R. Becker-Szendy {\it et al} Phys. Rev. {\bf D46}
(1992) 3720; D. Casper {\it et al} Phys. Rev. Lett. {\bf 66}
(1992) 2561; D. M. Roback, Measurement of the atmospheric
neutrino flavor ratio with Soudan 2, Ph D Thesis , Univ. of
Minnesota (1992); Ch. Berger {\it et al} Phys. Lett. {\bf B245}
(1990) 305; {\bf
B227} (1989) 489.
\bibitem{dm} A.N. Taylor and M. Rowan-Robinson, Nature {\bf
359} (1992) 396; M. Davis, F. J. Summers and D. Schlegel, Nature,
{\bf 359} (1992) 393.
\bibitem{moe} M.K. Moe, Talk at the Third Int. Workshop on
Theory and Phenomenology in Astroparticle and Underground
Physics, Gran Sasso, Italy (1993).
\bibitem{sanj} R. N. Mohapatra and G. Senjanovic, Phys. Rev.
{\bf D 23} (1981) 165.
\bibitem{msw} S.P. Mikheyev and A. Yu. Smirnov, Yad. Fiz. {\bf
42}  (1985) 1414; L. Wolfenstein, Phys. Rev. {\bf D17} (1978) 2369;
For a review on the msw effect, see, P.B. Pal, Oregon Univ. Preprint,
OITS 470 (1991).
\bibitem{cliff} P. Bamert and C. P. Burgess, Report No. McGill-94/07,
NEIP-94-003 (1994).
\bibitem{valle} A. Ioannissyan and J.W.F. Valle, Univ. of
Valencia preprint, FTUV/94-08 (1994).
\bibitem{moh1} D. Caldwell and Rabindra N. Mohapatra, Univ. of
Maryland report, UMD-PP-94-90 (1994).
\bibitem{moh2}D. G. Lee and R. N. Mohapatra, Univ. of Maryland Report,
UMD-PP-94-95 (1994).
\bibitem{lang}See for example, S. Bludman,D. Kennedy and P. Langacker,
 Nucl. Phys. {\bf B374} (1992) 373.
\bibitem{smir1} A. Yu. Smirnov, Phys. Rev. {\bf D48} (1993) 3264.
\bibitem{babu} R.N. Mohapatra and K.S. Babu,Phys. Rev. Lett.
{\bf 70}  (1993) 2845
\end{thebibliography}
\end{document}